\documentclass[]{spie}  %>>> use for US letter paper
\usepackage{amsmath,amsfonts,amssymb}
\usepackage{graphicx}
\usepackage[colorlinks=true, allcolors=blue]{hyperref}

\usepackage{amsfonts}
\usepackage{amsmath}
\usepackage{graphicx}      % include this line if your document contains figures
\usepackage{float} 
\usepackage{psfrag} 
\usepackage{mathtools}
\usepackage{url}
\usepackage{cite}

\title{A novel method for adaptive control of deformable mirrors}

\author{Aleksandar Haber}
\affil{Department of Manufacturing and Mechanical Engineering Technology, College of Engineering Technology, Rochester Institute of Technology,  70 Lomb Memorial Dr., Rochester NY 14623, USA}

\authorinfo{Further author information: E-mail: aleksandar.haber@gmail.com; awhmet@rit.edu, Telephone: +1 585-475-5813}

\begin{document} 
\maketitle

\begin{abstract}
For sufficiently wide ranges of applied control signals (control voltages), MEMS and piezoelectric Deformable Mirrors (DMs), exhibit nonlinear behavior. The nonlinear behavior manifests itself in nonlinear actuator couplings, nonlinear actuator deformation characteristics, and in the case of piezoelectric DMs, hysteresis. Furthermore, in a number of situations, DM behavior can change over time, and this requires a procedure for updating the DM models on the basis of the observed data. If not properly modeled and if not taken into account when designing control algorithms,  nonlinearities, and time-varying DM behavior, can significantly degrade the achievable closed-loop performance of Adaptive Optics (AO) systems. Widely used approaches for DM control are based on pre-estimated linear time-invariant DM models in the form of influence matrices. Often, these models are not being updated during system operation. Consequently, when the nonlinear DM behavior is excited by control signals with wide operating ranges, or when the DM behavior changes over time, the state-of-the-art DM control approaches relying upon linear control methods, might not be able to produce a satisfactory closed-loop performance of an AO system. Motivated by these key facts, we present a novel method for data-driven DM control. Our approach combines a simple open-loop control method with a recursive least squares method for dynamically updating the DM model. The DM model is constantly being updated on the basis of the dynamically changing DM operating points. That is, the proposed method updates both the control actions and the DM model during the system operation. We experimentally verify this approach on a Boston Micromachines MEMS DM with 140 actuators. Preliminary experimental results reported in this manuscript demonstrate good potential for using the developed method for DM control.
\end{abstract}


\section{INTRODUCTION}
\label{sec:intro}  % \label{} allows reference to this section
 To develop high-performance Adaptive Optics (AO) systems, it is of paramount importance to develop sufficiently accurate models of active optical components, such as Deformable Mirrors (DMs) or spatial light modulators~\cite{vogel2006modeling,stewart2007open,vogel2010modeling,haber2020modelingHaberVerhaegen}. Once high-fidelity models are established and properly validated, they can be used to develop high-performance model-based controllers~\cite{ruppel2012feedforward,mocci2020pi,kuiper2018advances,haber2016framework,manetti2013self,ravensbergen2009deformable,polo2013linear,bifano1999microelectromechanical,haber2013predictive,haber2013identification,kulcsar2012minimum,chiuso2009dynamic,song2011controller}. 

The main focus of this manuscript is on the development of control algorithms for DMs. In their essence, DMs are reflective optical elements whose surface deformations are precisely controlled by actuators. Widely-used DM actuation principles are based on MEMS, electromagnetic, electrostatic, piezoelectric, and ferrofluidic technologies and concepts. For sufficiently wide ranges of applied control actions (control voltages), MEMS and piezoelectric, as well as some other DM types, exhibit nonlinear behavior. The DM nonlinearities can manifest themselves in several forms. For example, for large magnitudes of applied control actions that are necessary to produce large mirror surface deformations, neighboring actuators become nonlinearly coupled through the observed DM surface deformation. This is especially the case for MEMS DMs~\cite{haber2021general}. Furthermore, a typical voltage-deformation response of a single actuator of a MEMS DM, is nonlinear~\cite{haber2021general}. On the other hand, DM concepts based on piezoelectric actuators are prone to hysteresis nonlinearities~\cite{ma2018hysteresis,schmerbauch2020influence}. In a number of scenarios and applications, DM behavior can change over time, and this requires online procedures for updating the DM models. For example, significant temperature fluctuations of the reflective surfaces, or temperature fluctuations of the main DM components, might significantly alter the DM behavior and dynamics. This is especially the case for space optics and for optics operating with high-power lasers, where  DMs can directly absorb a significant portion of external heat fluxes, or the heat conduction from the supporting elements and devices can cause nonuniform DM temperature increases and significant temperature gradients over active DM surface areas~\cite{haber2020modeling,haber2021modeling,haberThesis,haber2013identification,habets2016multi,saathof2015actuation,zhang2020optimization,banyal2013opto}. To properly analyze, predict, and control the influence of thermal phenomena on DM behavior, it is often necessary to use data-driven techniques to estimate thermal dynamics~\cite{haber2019identification,haber2020modeling,haber2019subspace}.

If not properly modeled and if not taken into account when designing control algorithms, these nonlinearities and time-varying DM behavior, can significantly degrade the achievable closed-loop performance of AO systems. Widely used approaches for DM control are based on pre-estimated linear time-invariant DM models in the form of influence matrices, see for example~\cite{Haber:13,haber2021general} and references therein. In addition to this, most of the control approaches do not update DM models during system operation. The widely-used linear control approaches might work satisfactorily well when the nonlinear DM behavior is not excited, and when the DM behavior is time-invariant~\cite{haber2021general}. For example, when the DM operating voltages are close to the bias voltages around which a linear model is estimated, we can expect that the linear DM model will be accurate, and consequently, a linear controller will work satisfactorily well. That is, we can expect that linear control methods will produce good results for small to medium voltage operating ranges. Since the achievable magnitudes of the Peak-to-Valleys (P-Vs) of DM surface deformations are directly correlated with the sizes of the voltage operating ranges,  linear control methods can relatively accurately correct for wavefront aberrations with small to medium P-Vs. However, if it is necessary to correct for wavefront aberrations with larger P-Vs, then the operating voltages need to significantly deviate from the linearization points. These deviations can significantly degrade the accuracy of linear models, and consequently, linear controllers might not be able to accurately correct for wavefront aberrations~\cite{haber2021general}. These key facts and observations motivate the development of novel controllers for DMs that will be able to extend the DM voltage operating ranges, as well as to successfully cope with DM nonlinearities and time-varying DM behavior. 

Nonlinear estimation and control approaches have received less attention in the AO community. The most probable reason for this is that compared to linear control methods, the design and implementation of nonlinear control methods are significantly more complex. There are several obstacles that need to be overcome to properly design and implement nonlinear DM control algorithms. One of the main obstacles is that it is challenging to postulate appropriate model structures that can accurately capture the nonlinear DM behavior. Too complex nonlinear models with a large number of parameters are impractical for real-time control. The method proposed in~\cite{guzman2010deformable} uses nonparametric estimation techniques to postulate and estimate DM models that to some degree can model the DM nonlinearities. However, in order to accurately fit DM models, this approach might require a large number of wavefront samples. In principle, machine learning and reinforcement learning techniques~\cite{landman2021self,ke2019self,nousiainen2021adaptive,haber2019steady,mashrafi2020machine} can be used to estimate nonlinear DM models and design control algorithms that can deal with nonlinearities. However, these approaches often need a large number of iterations to converge and reach the optimal correction performance. Speeding up the convergence rate by incorporating some forms of \textit{a priori} models is an important research topic. 

A viable alternative to purely nonlinear control approaches is to dynamically update linear model parameters on the basis of the actual values of DM control voltages and on the basis of the observed wavefronts. Gain scheduling techniques~\cite{leith2000survey} are based on the idea of defining a number of linear models and corresponding controllers, each of which is defined for a certain voltage operating range. On the basis of the most current values of the control voltages, gain scheduling techniques select the most appropriate model. Another approach is to use adaptive control methods~\cite{aastrom2013adaptive}. Adaptive control methods are dynamically updating model parameters on the basis of the observed wavefront aberrations. To the best of our knowledge, the adaptive control idea for DM control has received less attention, and the true potential of this method for DM control has not been thoroughly investigated. A simplified adaptive control method is used in~\cite{zou2009high} to iteratively calibrate a DM model. A recursive least-squares method~\cite{ljung1999} for updating DM models has been used in~\cite{huang2015high}. The limitation of the methods proposed in~\cite{zou2009high,huang2015high} is that they do not take into account the actuator limits. Furthermore, the adaptive control method proposed in~\cite{huang2015high} works completely in closed-loop, and consequently, special care needs to be dedicated to proper parameter tuning and the problems of persistency of excitation and estimation consistency~\cite{landau2011adaptive}. Recently, in~\cite{haber2021general,haber2021dual} we have proposed adaptive control algorithms for DM control. The approach presented in~\cite{haber2021general} is combining an open-loop control approach with a batch-data multivariable least-square problem for dynamically updating the DM influence matrix. This method does not update the DM model in the closed-loop, and its tuning is relatively simple. However, this method might require a large number of data samples to estimate the DM influence matrix. On the other hand, the method proposed in~\cite{haber2021dual}combines a recursive least-squares method with a feedback control algorithm. However, the tuning of this method is far from trivial and special attention needs to be dedicated to the problems of persistency of excitation and estimation consistency. 

In this manuscript, we investigate the possibility of combining a simple open-loop control method with a recursive least squares method for dynamical estimation of the DM influence matrix. The developed method explicitly takes into account actuator constraints, and thus it has significant advantages over the approaches proposed in~\cite{zou2009high,huang2015high}. In every iteration, the DM influence matrix is updated by using the recursive least-squares method, and the updated influence matrix is used to compute open-loop control actions. Consequently, since the control process is performed in  open loop and since only the parameters of the recursive least-squares method need to be tuned, the tuning of this method is simpler compared to the tuning of the adaptive feedback control approach proposed in~\cite{haber2021dual}. On the other hand, the advantage of the developed method over the method proposed in~\cite{haber2021general}, is that it requires fewer samples to converge. This is because the adaptive estimation of the influence matrix is performed by using the recursive least-squares method, instead of using the batch-data multivariable least-squares method. We experimentally verify the developed approach on a Boston Micromachines MEMS DM with 140 actuators. In this manuscript, we present preliminary control results, and further experimental investigation and improvements of the proposed approach will be presented in subsequent publications. The preliminary experimental results reported in this manuscript demonstrate good potential for using the developed method for DM control.

This manuscript is organized as follows. In Section~\ref{sec:controlAlgorithm}, we present the control method. In Section~\ref{sec:experimentalResults}, we present the experimental results. Conclusions and future research directions are briefly discussed in Section~\ref{sec:conclusions}.

\section{CONTROL METHOD}
\label{sec:controlAlgorithm}
In this section we present the control method. We assume the following DM model
\begin{align}
\mathbf{z}_{k+1}=L_{k}\mathbf{b}_{k}+\mathbf{d}_{k},
\label{controlEquation}
\end{align}
where $k$ is a discrete time instant at which the DM surface deformation is observed or control actions are sent to the DM ($k$ is also referred to as a control iteration in the sequel), $\mathbf{z}_{k} \in \mathbb{R}^{n}$ is a vector consisting of Zernike coefficients obtained by approximating the DM surface deformation by using Zernike basis functions, $n$ is a total number of Zernike coefficients, $L_{k}\in \mathbb{R}^{n\times m}$ is an influence matrix that can dynamically change at every iteration $k$, $\mathbf{d}_{k}\in \mathbb{R}^{n}$ is a vector modeling the effect of the measurement noise, and $\mathbf{b}_{k} \in \mathbb{R}^{m}$ is a vector of control inputs $u_{k,i}$, $i=1,2,3,\ldots, m$, constructed as follows
\begin{align}
\mathbf{b}_{k}=\begin{bmatrix}u_{k,1}^{\theta} & u_{k,2}^{\theta} & \ldots & u_{k,m}^{\theta}   \end{bmatrix}^{T},
\label{controlVectorB}
\end{align}
where $\theta\in \mathbb{R}$ quantifies the voltage-deformation nonlinearity of a single actuator, and $m$ is a total number of DM actuators. We use an experimental setup with a sensor that is able to directly obtain mirror surface deformations (for more details, see Section~\ref{sec:experimentalResults}). Consequently, the objective of the control method is to produce a desired mirror surface deformation that is expressed by using the Zernike basis functions. However, with minor modifications, the proposed method can be used if the control objective is to produce or correct general wavefronts that are expressed by using the Zernike basis functions or some other basis functions. 

To simplify the control algorithm implementation, we scale down the control inputs $u_{k,i}$ to the interval $[0,1]$, that is, $0\le u_{k,i } \le 1 $. The value of $\theta$ depends on the used mirror type. To test the developed method, we use a Boston Micromachines MEMS DM with 140 actuators. Often, it is assumed that $\theta=2$ for this DM type. However, in~\cite{haber2021general} we demonstrated that a more accurate estimate of this constant is $\theta= 1.742$. Consequently, we develop the control method by assuming $\theta= 1.742$. If other DM  types are used, then this constant should be re-estimated by using the procedure explained in~\cite{haber2021general}.
 
 To apply the recursive least squares method~\cite{ljung1999} for the estimation of the influence matrix, it is necessary to transform the equation \eqref{controlEquation} into a more suitable form. In this form, the entries of the influence matrix should appear as vector entries. For that purpose, we apply the $\text{vec}(\cdot)$ operator~\cite{verhaegen2007filtering} to \eqref{controlEquation}. When applied to a matrix, this operator produces a vector in which the influence matrix entries are stacked column-wise. By applying the $\text{vec}(\cdot)$ operator to \eqref{controlEquation}, we obtain 
\begin{align}
\mathbf{z}_{k+1}=H_{k}\mathbf{x}_{k}+\mathbf{d}_{k},
\label{vectorizedEq}
\end{align}

where $H_{k}\in \mathbb{R}^{n\times(n\cdot m)}$, $H_{k}=\mathbf{b}_{k}^{T} \otimes I_{n}$, $I_{n}\in \mathbb{R}^{n\times n}$ is $n\times n $ identity matrix, and $\mathbf{x}_{k}=\text{vec}\big(L_{k}\big)$, $\mathbf{x}_{k}\in \mathbb{R}^{n\cdot m}$ is a vector of influence matrix parameters, and the notation $\otimes$ denotes the Kronecker matrix product. For more information about vectorization of matrix equations and the Kronecker product see~\cite{laub2005matrix}.  
 
To initialize the recursive least-squares method, we need to have an initial value of the vector $\mathbf{x}_{k}$. That is, we need to estimate the initial values of the influence matrix parameters. We use a multivariable least-squares method explained in~\cite{haber2021dual,haber2021general,Haber:13} to estimate these parameters. For the time being, we assume that over a discrete-time horizon $k=0,1,2,\ldots,s-1$, of length $s\ge m$ ($s$ is a parameter selected by the user, however, it has to be larger than the number of DM actuators), the influence matrix $L_{k}$ is constant and equal to an initial value, denoted by $L_{0}$. Then, similarly to the approach used in~\cite{haber2021general}, on the basis of \eqref{controlEquation}, we can form a multivariable least-squares problem, and by solving it, we can obtain the initial value of the influence matrix parameters. This procedure produces the following estimate of the influence matrix
 \begin{align}
\hat{L}_{0}=ZB^{T}\Big(BB^{T} \Big)^{-1},
\label{initialValueEstimate}
\end{align}
where the notation $\hat{L}_{0}$ denotes an estimate of the initial value of the influence matrix, and  
\begin{align}
Z=\begin{bmatrix}\mathbf{z}_{1} & \mathbf{z}_{2} & \ldots & \mathbf{z}_{s}  \end{bmatrix}, \;\; B=\begin{bmatrix}  \mathbf{b}_{0} & \mathbf{b}_{1} & \ldots & \mathbf{b}_{s-1} \end{bmatrix},
\end{align}
and where $Z\in \mathbb{R}^{n\times s}$ and $B\in \mathbb{R}^{m\times s}$. The columns of $B$ are vectors $\mathbf{b}_{i}$, $i=0,1,\ldots, s-1$. Every vector $\mathbf{b}_{i}$ is constructed by randomly selecting the control input $u_{i,j}$, $j=1,2,\ldots, m$ from a normal distribution with the mean of $0.5$ and standard deviation of $0.15$. If a randomly generated value of $u_{i,j}$ is larger than $1$, then we set that value to $1$. Similarly, if the randomly generated value of $u_{i,j}$ is smaller than zero, we set that value to $0$. The matrix $Z$ consists of the vectors $\mathbf{z}_{i}$, $i=1,2,\ldots, s$, obtained by decomposing the observed DM surface deformation by using the Zernike basis functions.

The goal of the control algorithm is to produce a desired mirror surface shape. By expressing this desired surface shape in the Zernike basis, we define the vector of desired Zernike coefficients represented by the vector $\mathbf{z}_{D}$. By using the estimated influence matrix $\hat{L}_{0}$, we estimate the initial values of the control actions by solving the following open-loop optimization problem 
\begin{align}
& \min_{\mathbf{b}_{0}} \left\| \mathbf{z}_{D} - \hat{L}_{0}\mathbf{b}_{0}  \right\|_{2}^{2}, \label{optimizationProblem1}  \\
& \text{subject to:} \;\;\;  \underline{\mathbf{b}} \le \mathbf{b}_{0} \le \overline{\mathbf{b}}, \label{optimizationProblem2} 
\end{align}
where  $ \underline{\mathbf{b}}  \in \mathbb{R}^{m}$ and $\overline{\mathbf{b}} \in \mathbb{R}^{m}$ are the lower and upper bounds on the optimization variable $\mathbf{b}_{0}$, and the relation $\le$ is applied element-wise. Since the control inputs for the DM are scaled down to the interval $[0,1]$, we select  $\underline{\mathbf{b}}$ as a vector of zeros, and $\overline{\mathbf{b}}$ as a vector of ones. We solve the problem \eqref{optimizationProblem1}-\eqref{optimizationProblem2} by using the MATLAB function $\text{lsqlin}(\cdot)$. Let the solution of \eqref{optimizationProblem1}-\eqref{optimizationProblem2} be denoted by $\hat{\mathbf{b}}_{0}$. Once this solution is found, we can easily compute the control inputs $\hat{u}_{0,i}$ from the entries of $\hat{\mathbf{b}}_{0}$.  

Beside the initial value of the influence matrix parameters given by the vector $\hat{\mathbf{x}}_{0}=\text{vec}(\hat{L}_{0})$, and the initial  vector $\hat{\mathbf{b}}_{0}$ computed by solving \eqref{optimizationProblem1}-\eqref{optimizationProblem2}, we also need an additional matrix to initialize the control method presented below. For the initial iteration $k$ of the method presented below, we define the matrix $S_{0}=\delta I_{n\cdot m}$, $S_{0} \in \mathbb{R}^{(n\cdot m )\times (n\cdot m )}$, where $\delta >0 $ is a positive real parameter selected by the user, and $I_{n\cdot m}\in \mathbb{R}^{(n\cdot m) \times (n\cdot m)}$ is an identity matrix.
\\
\\
The control algorithm consists of the following two steps that are performed iteratively for $k=0,1,2,\ldots$

\begin{enumerate}

\item Observe the DM surface deformation and form the vector of Zernike coefficients $\mathbf{z}_{k+1}$. By using the values of $S_{k}$ and $\hat{\mathbf{x}}_{k}$ computed at the iteration $k$, update the influence matrix parameter vector $\hat{\mathbf{x}}_{k+1}$ by using the recursive least-squares method

\begin{align}
H_{k}&=\hat{\mathbf{b}}_{k}^{T} \otimes I_{n}, \label{recursiveComputation1} \\
F_{k+1}& =S_{k}H_{k}^{T}\Big(\beta I_{n} +H_{k}S_{k}H_{k}^{T} \Big)^{-1}, \label{recursiveComputation2}\\
S_{k+1}& =\frac{1}{\beta} S_{k} -\frac{1}{\beta} F_{k+1}H_{k}S_{k},  \label{recursiveComputation3} \\
\boldsymbol{\varepsilon}_{k+1}& = \mathbf{z}_{k+1}-H_{k}\hat{\mathbf{x}}_{k},  \label{recursiveComputation4} \\
\hat{\mathbf{x}}_{k+1}&=\hat{\mathbf{x}}_{k}+F_{k+1}\boldsymbol{\varepsilon}_{k+1}, \label{recursiveComputation5} 
\end{align}

where $0 < \beta \le 1 $ is a parameter selected by the user.
\item For the computed $\hat{\mathbf{x}}_{k+1}$, form the matrix $\hat{L}_{k+1}$ (by inverting the vectorization operator). Compute the control inputs by solving the optimization problem

\begin{align}
& \min_{\mathbf{b}_{k+1}} \left\| \mathbf{z}_{D} -\hat{L}_{k+1}\mathbf{b}_{k+1}  \right\|_{2}^{2}, \label{optimizationProblem1iterative}  \\
& \text{subject to:} \;\;\;  \underline{\mathbf{b}} \le \mathbf{b}_{k+1} \le \overline{\mathbf{b}}. \label{optimizationProblem2iterative} 
\end{align}

Once the solution $\hat{\mathbf{b}}_{k+1}$ is computed, form the control input vector $\hat{\mathbf{u}}_{k+1}$, apply the control inputs, and go to step 1. 
\end{enumerate}

We use  the MATLAB function $\text{lsqlin}(\cdot)$ to solve the problem \eqref{optimizationProblem1iterative}-\eqref{optimizationProblem2iterative}. The vector $\boldsymbol{\varepsilon}_{k+1}$, defined in \eqref{recursiveComputation4}, is called a model error, since this vector is the difference between the observed Zernike coefficients $\mathbf{z}_{k+1}$ and the model prediction given by $H_{k}\hat{\mathbf{x}}_{k}$.

\section{Experimental Results}
\label{sec:experimentalResults}

In this section, we present experimental results obtained by testing the developed control method. The experimental setup for testing the developed control method is the same as the experimental setup used in~\cite{haber2021general}. Consequently, we only briefly describe the experimental setup. To test the control method we use a Boston Micromachines MEMS DM. The mirror surface is deformed by 140 actuators distributed over a 12 by 12 actuation grid with $4$ inactive corner actuators. The mirror has a pitch of $400$ $[\mu m ]$ and a stroke of about $2$ $[\mu m ]$. For more information about the used DM, see~\cite{haber2021general,haber2021dual} and references therein. The mirror surface shape is observed by using the Partitioned Aperture Wavefront (PAW) sensor~\cite{parthasarathy2012quantitative,barankov2013single}. The light source is an LED ($660$ $[nm]$, Thorlabs). 

The control algorithm is implemented in MATLAB.  The surface deformation (surface profile) observed by the PAW sensor is represented by an image of $1001$ by $999$ pixels. However, this area also covers an inactive mirror surface area. In our experiments, a central circular region of this image is used as an observation area and for approximating the mirror surface deformation by using the Zernike basis coefficients. The diameter of the circular region is $398$ pixels. A small percentage of actuators is outside of this observation area. However, the deformation caused by these boundary actuators can be observed in the circular observation region.

In every control iteration, we quantify the control method accuracy by computing the Root-Mean-Square (RMS) value of the surface shape error. The surface shape error is computed by subtracting the desired surface shape from the produced surface shape.

\begin{figure}[H]
\centering 
\includegraphics[scale=0.6,trim=0mm 0mm 0mm 0mm ,clip=true]{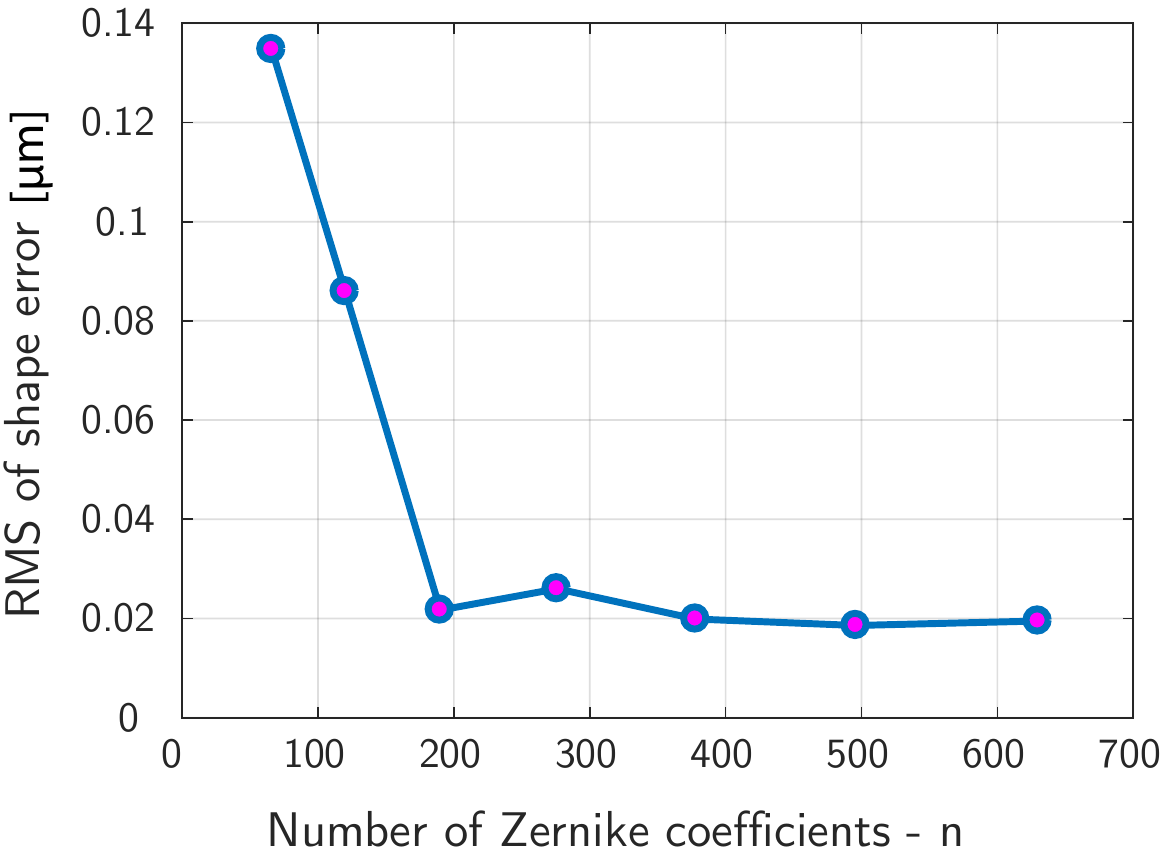}
\caption{RMS of surface shape errors as the function of the number of the Zernike coefficients (represented by $n$) that are used to approximate the mirror surface deformation.}
\label{fig:Graph1}
\end{figure}

To properly implement the control method it is important to select a sufficiently large number of Zernike coefficients (represented by $n$) for approximating the mirror surface shape. Namely, our procedure for estimating the initial value of the influence matrix is based on random control inputs for DM actuators. These random control inputs excite higher-order spatial frequencies of the mirror surface shape. Consequently, it is necessary to select a sufficiently large value of $n$ to accurately approximate the mirror surface shape by using the Zernike basis decomposition. Small values of $n$ will produce influence matrix models with significant model uncertainties. These model uncertainties can significantly limit the control accuracy or even cause control instabilities. To illustrate this effect, we vary $n$, and for every value of $n$ we estimate the initial value of the influence matrix by computing \eqref{initialValueEstimate}. For every value of $n$, we use $s=400$ in \eqref{initialValueEstimate} ($s$ is the number of data samples used to estimate the influence matrix). The desired mirror surface shape is equal to a scaled and shifted version of $Z_{4}^{2}$ (vertical secondary astigmatism). For this desired shape, we compute the initial control inputs by solving \eqref{optimizationProblem1}-\eqref{optimizationProblem2}. We then initialize the iterative control method with the initial value of the estimated influence matrix and the initial value of the computed control inputs. We use the following parameters $\beta=0.98$ and $\delta=10^{-2}$ in the control algorithm. After $30$ iterations of the control algorithm, we identify the smallest value of the RMS shape error. While computing the RMS value, we crop the edges of the circular domain over which the surface shape error is defined to exclude the edge effects (the edge effects are illustrated later in the text). The results are shown in Fig.~\ref{fig:Graph1}. 

The large RMS values of the surface shape errors for smaller values of $n$ are caused by the Zernike basis decomposition errors. For smaller values of $n$, we are not able to accurately decompose the mirror surface shape. Consequently, the estimated influence matrices are not able to accurately represent the DM behavior. On the other hand, we can observe that after approximately $n=370$, the RMS values of the surface shape errors saturate. That is, by using the values of $n$ larger than $370$, we do not obtain any additional gains in control accuracy. This limit is important to know since larger values of $n$ significantly increase the computational and memory complexities of implementing the recursive least-squares method \eqref{recursiveComputation1}-\eqref{recursiveComputation5}. By excessively increasing $n$, we will not improve the control accuracy, while on the other hand, we will significantly increase the computational and memory complexities.

\begin{figure}[H]
\centering 
\includegraphics[scale=0.6,trim=0mm 0mm 0mm 0mm ,clip=true]{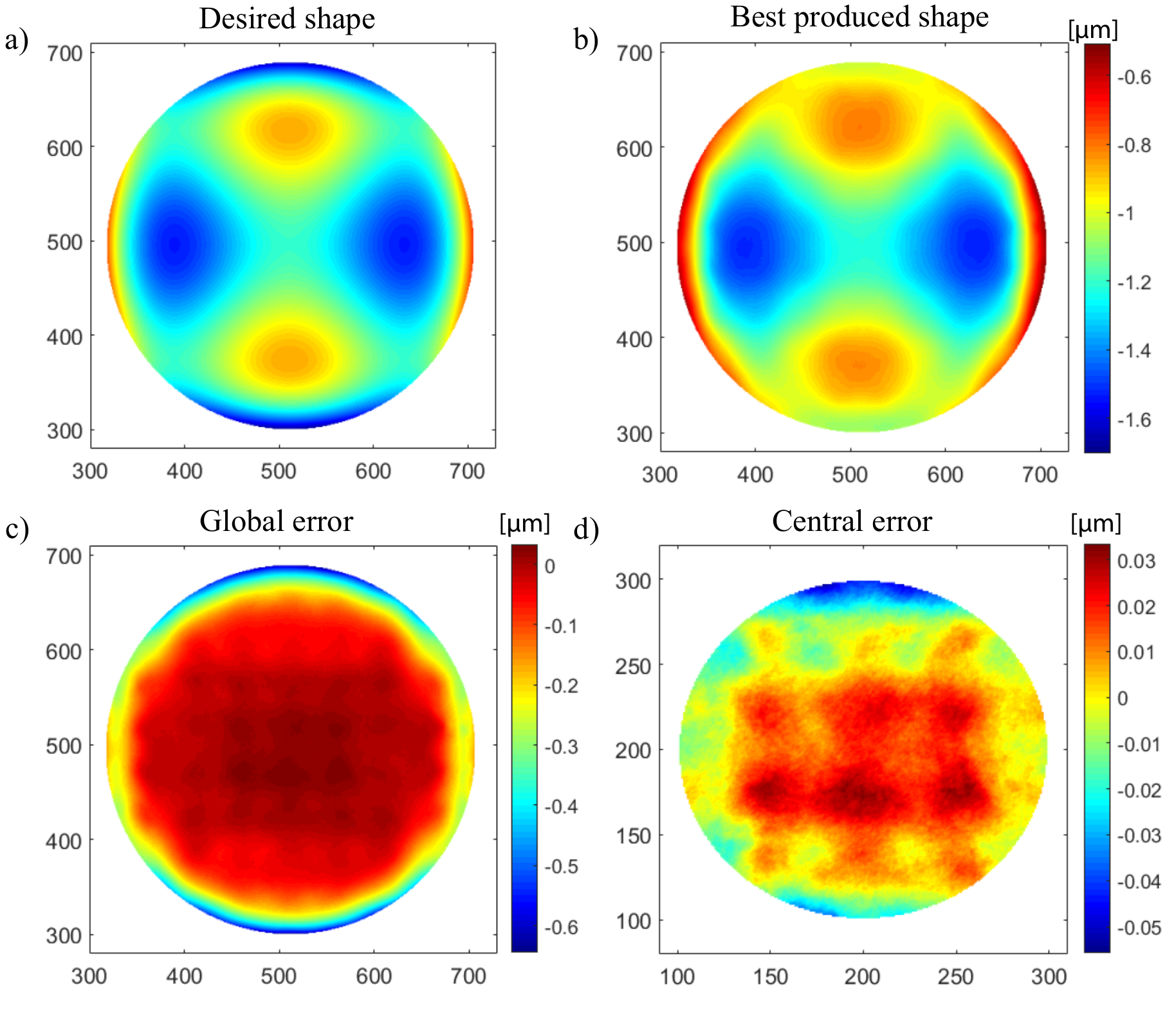}
\caption{Generation of the desired shape that is equal to a scaled and shifted version of $Z_{4}^{2}$. (a) Desired surface shape. (b) Best produced shape. (c) Global surface shape error. (d) Central surface shape error after edge effects are cropped from the global surface shape error shown in panel (c). The RMS value of the central surface shape error is $0.0186$ $[\mu m]$.}
\label{fig:Graph2}
\end{figure}

Next, we investigate the accuracy and the convergence of the developed method. First, we test the method for the desired surface shape equal to a scaled and shifted version of $Z_{4}^{2}$, and for the control algorithm parameters $\beta=0.98$ and $\delta=10^{-2}$. To approximate the mirror surface deformation we use $n=498$ Zernike coefficients. The P-V of the desired surface shape is $1.1829$ $[\mu m]$. For the estimation of the initial value of the influence matrix, we use $s=400$ data samples. After generating the initial guesses of the control inputs and the influence matrix, we initialize and start the control method. The results are shown in Fig.~\ref{fig:Graph2}. Panel (a) shows the desired surface shape. Panel (b) shows the best-produced shape. This shape is produced at iteration $9$ of the developed method. Panel (c) shows the global surface shape error corresponding to the best-produced shape. We can notice that the close to the edges, the error values significantly increase. Panel (d) is obtained by cropping the surface shape error such that the edge effects are removed. This is the central surface shape error. The RMS value of the central surface shape error is $0.0186$ $[\mu m]$.

\begin{figure}[H]
\centering 
\includegraphics[scale=0.75,trim=0mm 0mm 0mm 0mm ,clip=true]{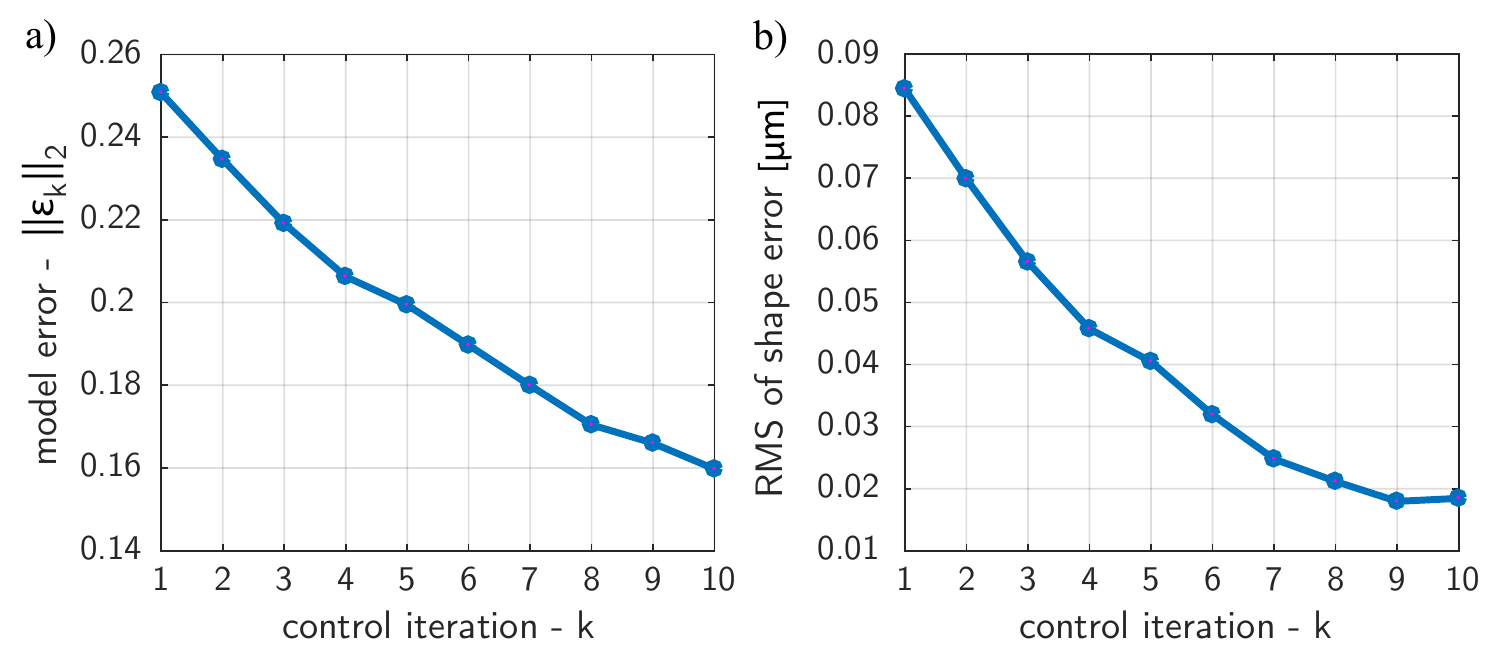}
\caption{Convergence of the developed control method. (a) Convergence of the model error quantified by the 2-norm $\left\|\boldsymbol{\varepsilon}_{k} \right\|_{2}$. Convergence of the RMS of the (central) shape error.}
\label{fig:Graph3}
\end{figure}

\begin{figure}[H]
\centering 
\includegraphics[scale=0.6,trim=0mm 0mm 0mm 0mm ,clip=true]{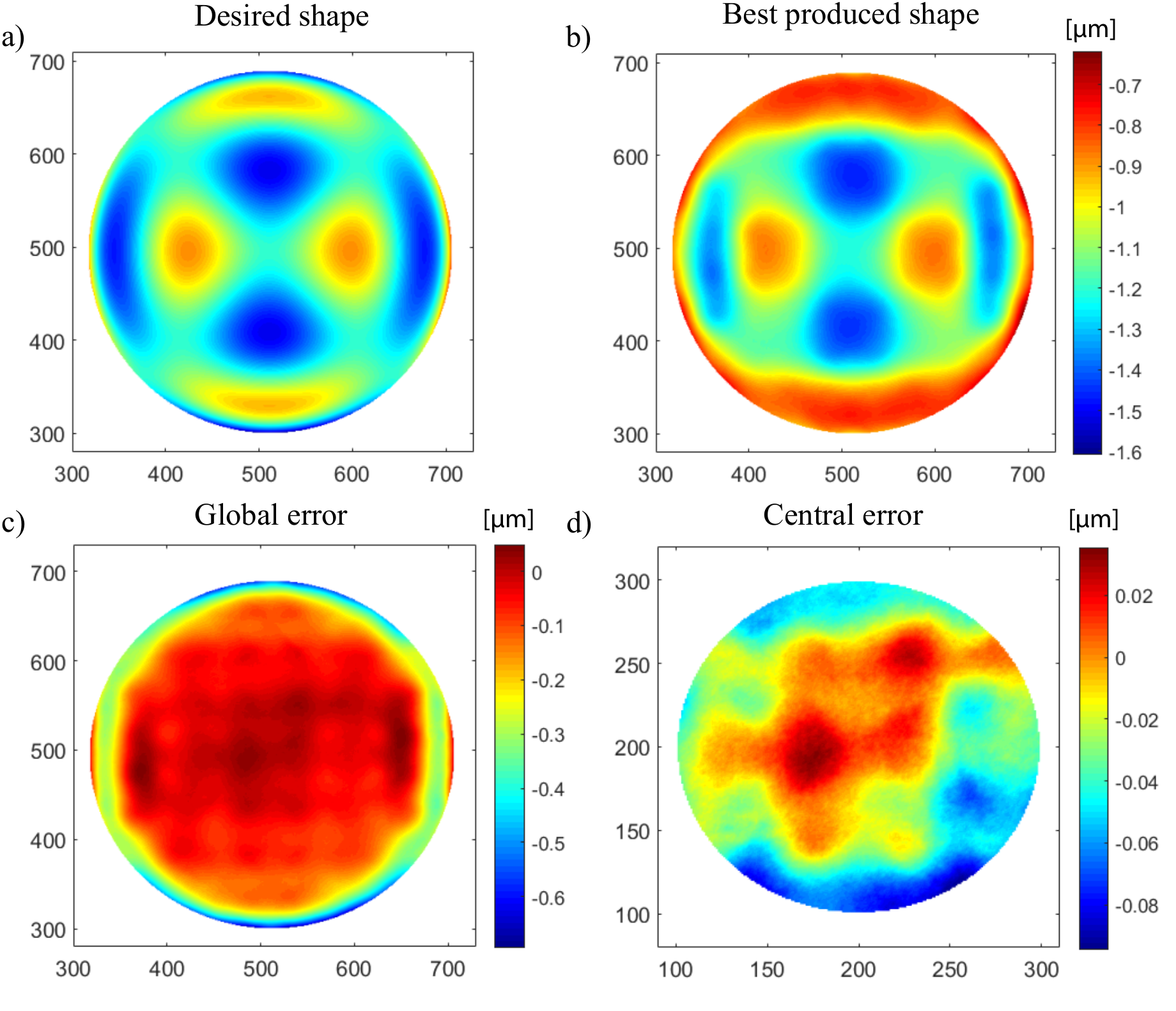}
\caption{Generation of the desired shape that is equal to a scaled and shifted version of $Z_{6}^{2}$. (a) Desired surface shape. (b) Best produced shape. (c) Global surface shape error. (d) Central surface shape error after edge effects are cropped from the global surface shape error shown in panel (c). The RMS value of the central surface shape error is $0.0393$ $[\mu m]$.}
\label{fig:Graph4}
\end{figure}

Figure~\ref{fig:Graph3} shows the convergence of the control method for the desired shape used to generate the results shown in Fig.~\ref{fig:Graph2}. The convergence is quantified by computing the 2-norm of the model error vector $\boldsymbol{\varepsilon}_{k+1}$ defined in \eqref{recursiveComputation4}, and by computing the RMS of the central shape error with the domain shown in Fig.~\ref{fig:Graph2}(d). 
Figure~\ref{fig:Graph4} shows the control performance for producing a shifted and scaled version of $Z_{6}^{2}$. The P-V of the desired shape is $0.9897$ $[\mu m]$. The control parameters are $\beta=0.98$ and $\delta =10^{-6}$. For the estimation of the initial value of the influence matrix, we use $s=400$ data samples.  The RMS value of the central surface shape error is $0.0393$ $[\mu m]$.  

From Figs.~\ref{fig:Graph2} and \ref{fig:Graph4} we can observe that the values of the surface shape error increase close to the edges. There are two approaches for reducing these errors. The first approach is to decrease the diameter of the circular domain area over which Zernike decomposition is defined. In this way, we will have more boundary actuators (some of which will be outside the circular domain area, however, their effect will still be observable) that will be able to control the deformation close to the boundaries. The second approach is to tune the parameters of the control algorithms. The main parameters are $\delta$, $s$, and $\beta$. Optimal tuning of the control parameters is left for future research. 

As mentioned previously, the computational complexity of the developed approach significantly increases with the number of Zernike coefficients used to approximate the mirror surface deformation. Furthermore, the computational complexity also increases with the number of actuators of the DM. The main computational bottleneck originates from the recursive least-squares method given by the equations \eqref{recursiveComputation1}-\eqref{recursiveComputation5}. One of the pathways for reducing the computational complexity is to exploit the Kronecker structure and hidden sparsity structure of the problem. The starting point for the research activities directed toward reducing the computational complexity can be approaches presented in~\cite{massioni2011fast,haber2016framework,haber2018sparsity,cerqueira2021sparse,sinquin2018tensor,haber2014subspace,massioni2015approximation,Haber:13mhe,haber2012identification}.

\section{Conclusion and Future Work}
\label{sec:conclusions}

In this manuscript, we developed a novel method for the adaptive control of Deformable Mirrors (DMs). The developed method relies on a recursive least-squares method for updating the influence matrix of the DM, and on an open-loop control method for controlling the DM. The developed method is experimentally verified by using a Boston Micromachines MEMS DM with 140 actuators. Preliminary experimental results reported in this manuscript demonstrate good potential for using the developed method for DM control. The future research direction should be directed toward developing an approach for tuning the parameters of the developed method. Also, research efforts should be directed toward reducing the computational and memory complexities of the developed approach.

\acknowledgments % equivalent to \section*{ACKNOWLEDGMENTS}

We would like to thank Professor Thomas Bifano from Boston University for enabling us to use the Boston Micromachines MEMS DM.

% References
\bibliography{sample} % bibliography data in report.bib
\bibliographystyle{spiebib} % makes bibtex use spiebib.bst

\end{document}